\documentclass{cjaa}                    

\usepackage{graphicx}                   
\input{epsf.sty}                        
\input{psfig.sty}                       

\begin{document}

\newcommand{\itbf}[1]{\textbf{\textit{#1}}}

\title{\itbf{INTEGRAL} and \itbf{Swift}  Observations of Blazars in Outburst
}


   \author{Elena Pian
      \inst{1}\mailto{pian@oats.inaf.it}
   \and Luigi Foschini
      \inst{2}
   \and Gabriele Ghisellini
      \inst{3}
      }
   \offprints{E. Pian}                   

   \institute{INAF, Trieste Astronomical Observatory,
             Via G.B. Tiepolo 11, I-34143 Trieste, Italy\\
             \email{pian@oats.inaf.it}
        \and
            INAF-IASF, Via P. Gobetti 101, I-40129 Bologna, Italy \\
        \and
            INAF, Brera Astronomical Observatory, Via E. Bianchi 46, I-23807 Merate (LC), Italy\\
          }

   \date{Received~~2007 month day; accepted~~2007~~month day}

   \abstract{The blazars 3C~454.3, PKS~0537-441 and PKS~2155-304 are traditionally known to be among the most active sources of this class.  They emit at all frequencies, up to the gamma-rays, and are good probes of multiwavelength nuclear variability.  The first two have also luminous broad emission line regions.  We have recently monitored them with various facilities, including {\it Swift} and {\it INTEGRAL}, and have interpreted their variations with models of non-thermal radiation from a relativistic jet.  In particular, we have tested for the first two sources the hypothesis that the variability is produced within the jet through internal shocks, i.e. collisions of relativistic plasma blobs. This allows a parameterization of all physical quantities as functions of the bulk Lorentz factor.  We have made the critical assumption that  every flaring episode is characterized by a fixed amount of energy.  The model reproduces brilliantly the multiwavelength data,  and especially the gamma-ray spectra, when available.  The model is not applicable to PKS~2155-304, the variability of which is caused by independent variations of few individual parameters.
}

   \authorrunning{E. Pian, L. Foschini \& G. Ghisellini}            
   \titlerunning{Observations of Blazars}  


   \maketitle
%
%
\section{Introduction}           
\label{sect:intro}
Relativistic jets are detected in both Galactic and extragalactic sources.  In the latter case, they reach kiloparsec, or even Megaparsec, scales and produce different observed phenomena, radiogalaxies and blazars, according to whether they are viewed sideways or end-on (Urry \& Padovani 1995).  Within this unified scheme, blazars are therefore ideal probes of  extragalactic jets,  because their orientation influences - through relativistic aberration - the jet kinematics and enhances the emission variability at all wavelengths. 

Ambitious multiwavelength campaigns have been organized in the last 10 years on selected blazar sources, to monitor the variability of the whole spectrum in different emission states and on different time scales, to identify correlated variations at various frequencies, and to constrain the models (Ulrich, Maraschi \&
Urry 1997; Pian et al. 1998; Tagliaferri et al. 2003; 
Krawczynski et al. 2004; Dermer \& Atoyan 2004;
B{\l}a\.zejowski et al.
2005; B\"ottcher et al. 2005; Sokolov \& Marscher 2005; Aharonian et 
al. 2006; Albert et al. 2006; Kato, Kusunose \& Takahara 2006; Massaro et al. 2006;
Raiteri et al. 2006). 

While the mechanism by which the inner engine (a supermassive, possibly rotating,  black hole) converts gravitational into kinetic energy and transfers it to the relativistic plasma is still unknown and the problems related to the exact interplay between the compact central object, its surrounding disk and the jet are still to be solved (Maraschi \& Tavecchio 2003; Vlahakis \& K\"onigl 2004; McKinney 2006), a clear paradigm has emerged for the production of the multiwavelength energy distributions of blazars: it is commonly accepted that synchrotron radiation dominates the spectrum from the radio to the UV (and occasionally X-ray) domain, while inverse Compton scattering prevails at higher energies.  The radiating plasma is accelerated within the jet, and propagates relativistically through disturbances and shocks, which are responsible for the variability.  The role of components external to the jet, such as the accretion 
disk/torus and the broad emission line region (BLR) has been recognized to be critical in the spectrum formation, and particularly in providing seed photons for the inverse Compton scattering (external Compton, 
Dermer \& Schlickeiser 1993; 
Sikora, Begelman, \& Rees 1994; Ghisellini \& Madau 1996; Ghisellini et al. 1998;
B{\l}a\.zejowski et al. 2000; 
Celotti, Ghisellini
\& Fabian 2007).

When these components are bright and relevant with respect to the jet emission, the "external" contribution to the inverse Compton scattering process becomes significant, or even dominant, with respect to the "internal" synchrotron self-Compton process (i.e., inverse Compton of the relativistic particles off the jet synchrotron photons) and generates differences in the broad-band spectra.  The differences among the different blazar "flavours" (Flat Spectrum Radio Quasars, Low-Energy Peaked BL Lacs, High-Energy Peaked BL Lacs) can be explained by differences in the relative importance of the Compton cooling and therefore, ultimately, by the different role of the BLR, powered by the thermal accretion disk (Ghisellini et al. 1998; Ghisellini, Celotti, \& Costamante 2002).

Our recent multiwavelength observing campaigns of blazars benefitted from the joint availability of high energy facilities ({\it INTEGRAL}, {\it Swift}) and ground-based flexible small optical/infrared monitors, like the Rapid Eye Mount (REM).  They were triggered and driven by the detection of an outburst and were aimed, through the comparison of low and high emission states, at determining the parameters responsible for variability, and the role played by the BLR photon reservoir during different states.
The  "economic" jet model (\S 2), applied to a scenario of internal shocks in the blazar jet, provides a physically meaningful description of some  observations.  We report here our tests of the economic model on two sources with strong emission lines, 3C~454.3 and PKS~0537-441, and compare their variability with that of the "classical" 
featureless BL Lac object PKS~2155-304 (\S 3).   We discuss our results and the model applicability in \S 4.

\section{An "economic" jet model}
\label{sect:Theo}

The jet model we adopt has been presented in Katarzi\'nsky and Ghisellini (2007) and is based on the
scenario of internal shocks popularly applied to Gamma-Ray Bursts (M\'esz\'aros \& Rees 1994; Sari \& Piran 1997), but
originally developed for extragalactic kiloparsec jets (Rees 1978).  Applications of the internal shock scenario to individual classes of blazar sources has been presented in Spada et al. (2001) and Guetta et al. (2004).   Relativistic plasma blobs of different velocity collide within the jet (internal shocks), merge into a single blob and give rise to the observed multiwavelength outbursts.   Direct evidence of jet components traveling at different velocities has been provided by the VLBI radio measurements (Abraham et al. 1996; Jorstad et al. 2001a), although these map the jets on  several parsec scales, much larger than those where the outbursts take place, that are  at most few light-days across, as inferred from emission variability. 

The basic assumptions of  the model are 1) that the jet has a fixed efficiency, i.e. each blob receives the same amount of energy  from the central engine, so that its maximum Lorentz factor is inversely proportional  to its mass, and 2)  that the contrast in the Lorentz factors of the colliding shells is always the same, i.e. the amount of energy transmitted to the emitting electrons during a collision is constant.   In the internal shock model, the distance at which the blobs collision occurs from the jet apex is proportional to the square of the lower Lorentz factor.   Therefore, slower blobs collide closer to the nucleus than faster blobs.   

Since all physical quantities scale with the distance from the nucleus, the site of the collision and dissipation  is critical for determining the dominance of a radiation component over the other.  Close to the nucleus, the magnetic field experienced by the plasma is stronger, and the influence of the BLR is weaker; this suggests a more significant synchrotron emission with respect to external Compton.  The opposite is true when the collision occurs farther from the nucleus: the magnetic field has lower strength and the BLR photon density is larger.   Thus, at different sites along the jet, the synchrotron and inverse Compton (primarily external 
Compton) components may have different ratios, even if the injected total energy is the same.

\section{The observing campaigns}
\label{sect:data}

\subsection{\itbf{INTEGRAL} observations of 3C~454.3}

We activated our {\it INTEGRAL} program for observations of blazars in outburst in May 2005, following the dissemination of an optical alert for the Flat Spectrum Radio Quasar 3C~454.3 ($z = 0.859$).   The source was in an unusually bright state ($V \sim 12$), and we verified that also the RXTE All Sky Monitor was registering a period of X-ray activity.   Many orbiting and ground-based observatories started a monitoring (Giommi et al. 2006; Fuhrmann et al. 2006; Pian et al. 2006; Villata et al. 2007). The spectral energy distributions of the blazar  in Spring 2005, based on {\it INTEGRAL} data, and at  previous epochs are reported in Figure 1, along with a sketch of the jet model used to reproduce the two different multiwavelength states.  

The first 2 blobs ($\Gamma_1$ and $\Gamma_2$) are fast and collide far from the central engine, but within the BLR, that is known to be a relevant source of photons  in 3C~454.3 (Pian et al. 2005).  Therefore, the   Compton scattering of the jet electrons off the BLR photons is very significant,
because the density of the external radiation in the rest frame of the blobs is high.  This external Compton component peaks in the MeV-GeV range, and the model indeed matches well the EGRET spectrum of this blazar observed in 1991-1994.  In Spring 2005, the blobs collision occurs closer to the center, so that the synchrotron component is enhanced with respect to inverse Compton.  The difference in the initial Lorentz factors of the slower blobs at the two epochs is less than a factor 2: the Lorentz factor corresponding to the  "historical" state   is $\Gamma = 11$, and that of May 2005 is $\Gamma = 6.25$.  In Figure 1 are also shown the synthetic spectra obtained for a range of Lorentz factors in between these two values.

\begin{figure*}
  \vspace{2mm}
   \begin{center}
   \hspace{3mm}\psfig{figure=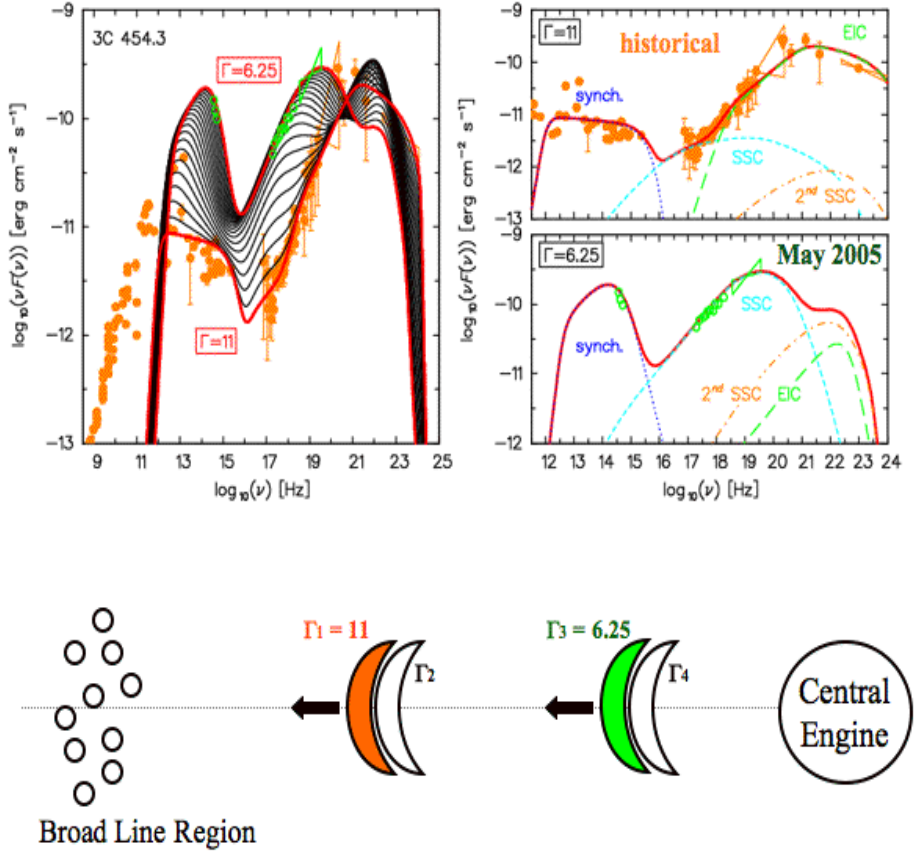,width=150mm,height=130mm,angle=0.0}
   \parbox{180mm}{{\vspace{2mm} }}
   \caption{Multiwavelength energy distributions of the blazar 3C~454.3 (top) and scheme of a relativistic "jet" (bottom, not to scale): a  blob of Lorentz factor $\Gamma_1 = 11$ (colored in orange)  is ejected at a certain time   from the central engine.  A following faster blob with
$\Gamma_2 > \Gamma_1$  collides with it and produces an outburst, the spectrum of which is reported in orange in the two top panels ("historical" state).  At a subsequent epoch, a blob of Lorentz factor 
$\Gamma_3 = 6.25$ (green) is ejected, and is hit by a following blob of $\Gamma_4 > \Gamma_3$ ejected soon thereafter.  This collision produces the outburst spectrum reported in green in the above panels (state of May 2005).  The differences between the two multiwavelength spectra is completely accounted for by the difference of the bulk Lorentz factors at the 2 epochs (see text).  A family of model spectra, parameterized by the Lorentz factor (the step is $\Delta\Gamma = 0.25$), is shown in the top left panel: the importance of the external inverse Compton component  increases with the Lorentz factor and the dominance of the synchrotron component decreases accordingly. See data references in Pian et al. (2006) and more model details in Katarzy\'nsky \& Ghisellini (2007).}
   \end{center}
\end{figure*}

\subsection{ \itbf{Swift} observations of PKS~0537-441}

This blazar ($z = 0.896$)  has been observed at various epochs at many wavelengths and is known for its remarkable variability  (see Pian et al. 2007, and references therein).  Like 3C~454.3, it has a luminous BLR (Pian et al. 2005).  In 2005 it has been monitored in the optical and infrared by REM (Dolcini et al. 2005) and observed by all instruments of {\it Swift} in January, July and November.  Figure 2 reports the XRT light curve in 2 energy bands and the optical light curve in V-band obtained by the combination of the UVOT and REM observations.  The X-ray light curves show a remarkable flare (factor of $\sim$4), but the simultaneous optical variations are astonishing:  the source increased by a factor of $\sim$60  over about 1 month between December 2004 and January 2005, and then decreased during 2005.  

\begin{figure}
  \vspace{2mm}
   \begin{center}
   \hspace{3mm}\psfig{figure=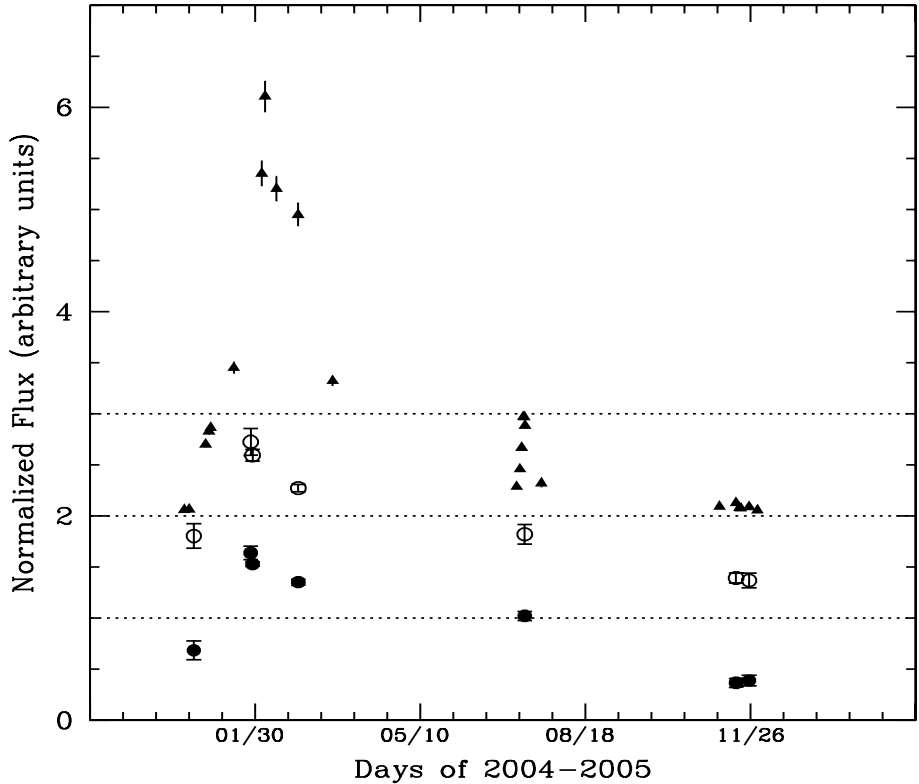,width=150mm,height=130mm,angle=0.0}
   \parbox{180mm}{{\vspace{2mm} }}
   \caption{{\it Swift}/XRT  background-subtracted light  curves of PKS~0537--441 in the  
1-10 keV (filled circles) and in the 0.2--1\,keV (open circles) 
energy bands, and optical
light  curve (triangles),  obtained from  the  merging of  the UVOT  V
filter and REM V filter observations. 
The curves are not corrected for
Galactic extinction,  and are normalized to  their respective averages
(0.136 cts~s$^{-1}$  in the 1-10  keV band, 0.084 cts~s$^{-1}$  in the
0.2-1 keV band, 6.58 mJy in  the optical band).
The dotted horizontal lines indicate the
average values of the three light curves: for clarity, the 0.2-1
keV  band and  V-band light  curves have  been scaled  up  by additive
constants 1 and 2, respectively.  Note that this upscaling implies that the
flux ratios derived by direct inspection of the soft X-ray (0.2-1 keV) and
optical light curves
do not correspond to the real 
ones, the fluxes having been increased by constants 1 and 2, respectively. 
The maximum amplitudes of variability in optical and X-rays are a factor of $\sim$4 and $\sim$60, respectively (from Pian et al. 2007).}
   \end{center}
\end{figure}

We have constructed the spectral energy distributions of the blazar using the simultaneous {\it Swift} and REM data of our campaign , and have compared them to the historical multiwavelength spectra.  The collection of the 2005 energy distributions and the two historical ones
are shown in the left and right panel of Figure 3, respectively.   We have modeled all multiwavelength spectra with the Katarzy\'nski \& Ghisellini (2007) model, by accounting for the multiwavelength variability only with variations of the bulk Lorentz factor $\Gamma$ and by parameterizing every other physical quantity as a function of $\Gamma$.   The fitting curves are very satisfactorily reproducing the data. The variability is due to rather small variations of $\Gamma$: from a minimum of  $\Gamma \simeq 10$ in the most luminous state of February 2005 to a maximum of  $\Gamma \simeq 15$  in the dimmest states of November 2005 as well as in the low states prior to 2005.  The intermediate state of July 2005 is accordingly described by  $\Gamma \simeq 12$.   
  
Some  physical  quantities yielded by  the model are reported in Figure 4: the total
luminosities associated with the protons, electrons and magnetic fields have a very weak or null dependence from the Lorentz factor (see lower panel of Fig. 4), indicating that the bolometric energy input is constant at all epochs.  The spectral differences are related to the location of the dissipation site along the jet.  Note that the MeV-GeV flux observed by EGRET for this blazar   in 1991-1992 and in 1995 is well reproduced by the model curves.  Therefore, it would have been crucial to observe the PKS~0537-441 at these energies simultaneously with the X-ray and optical observations, because the model predicts here a large variability.  

   \begin{figure}
   \plottwo{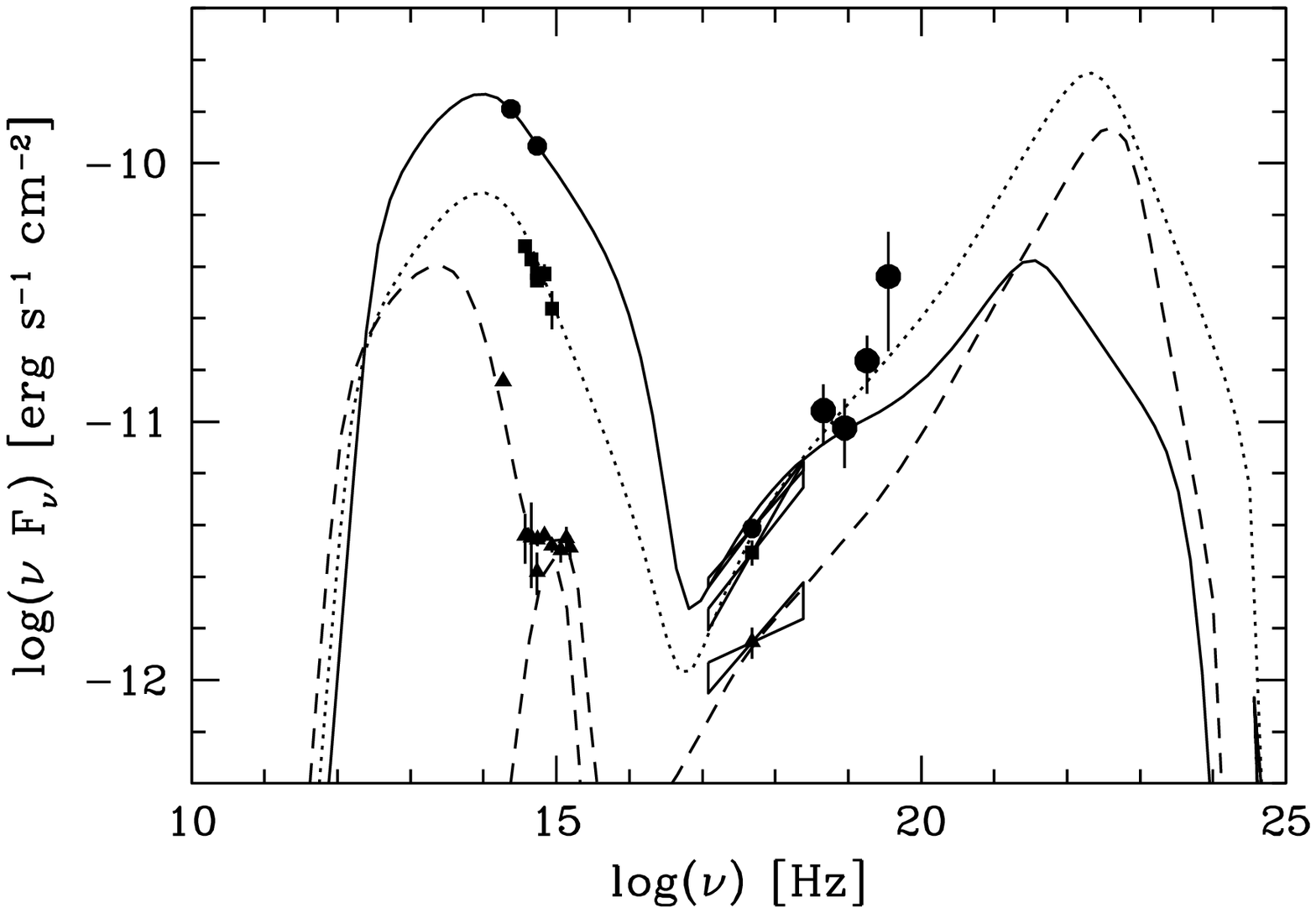} {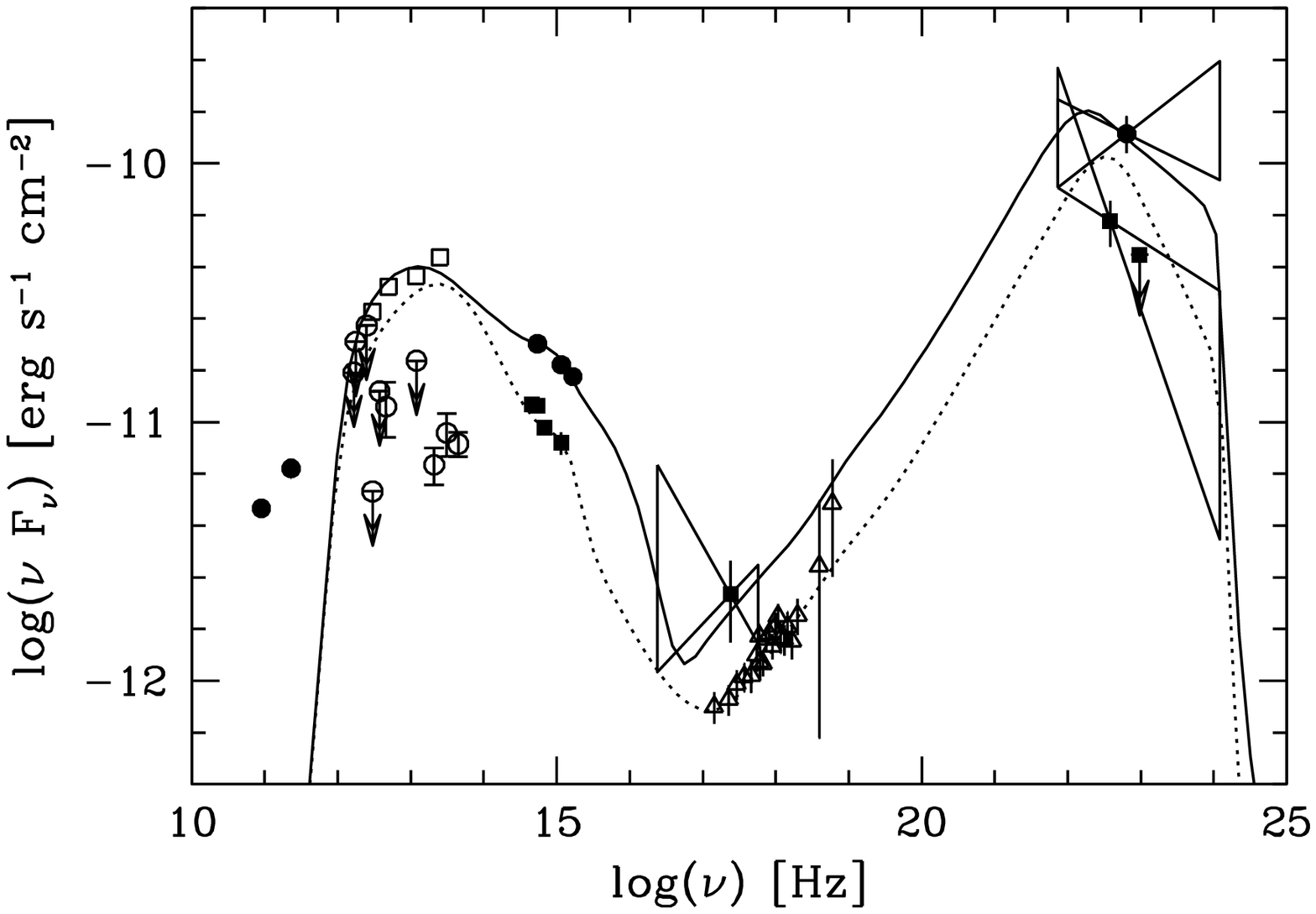}
   \caption{Spectral energy distributions of PKS~0537-441. {\it   Left}: The multiwavelength spectra refer to  24-25 February
2005 (small circles),  12 July 2005 
(squares) and 24 November  2005 (triangles).  The big circles represent the {\it Swift} BAT data. The 
{\it Swift} XRT data are reported along with the 1 $\sigma$ confidence ranges
of their power-law  fits.   
The flux uncertainties are  1 $\sigma$ (in some cases they are smaller than the
symbol size).
The  X-ray, UV,
optical and  near-IR data are  corrected for Galactic  extinction (see Pian et al. 2007).   Overplotted are the jet models (Katarzy\'nski \& Ghisellini 2007,
see text) for the energy distributions of 24-25 February 2005
(solid curve), 12 July 2005 (dotted curve), 24 November 2005 (dashed curve). The
thermal component required to account for the observed optical-UV
flux is also reported as a dashed curve.
{\it Right}:  Spectral energy distributions of PKS~0537-441  in 1991-1992 (filled squares) and 1995 (filled circles).   The 1 $\sigma$ confidence ranges
of the EGRET spectra are reported as light dashed lines.
The far-infrared data taken
by IRAS and ISO and the X-ray BeppoSAX data are not simultaneous and are
represented as open squares, open circles and open triangles, respectively (see
Pian et al. 2002, and references therein; Padovani et al. 2006). 
These spectra  have been also  
modelled according to Katarzy\'nski \& Ghisellini (2007): the model curves for
the 1991-1992 and 1995 states are dotted and solid, respectively (from Pian et al. 2007).}
\end{figure}

\begin{figure}
  \vspace{2mm}
   \begin{center}
   \hspace{3mm}\psfig{figure=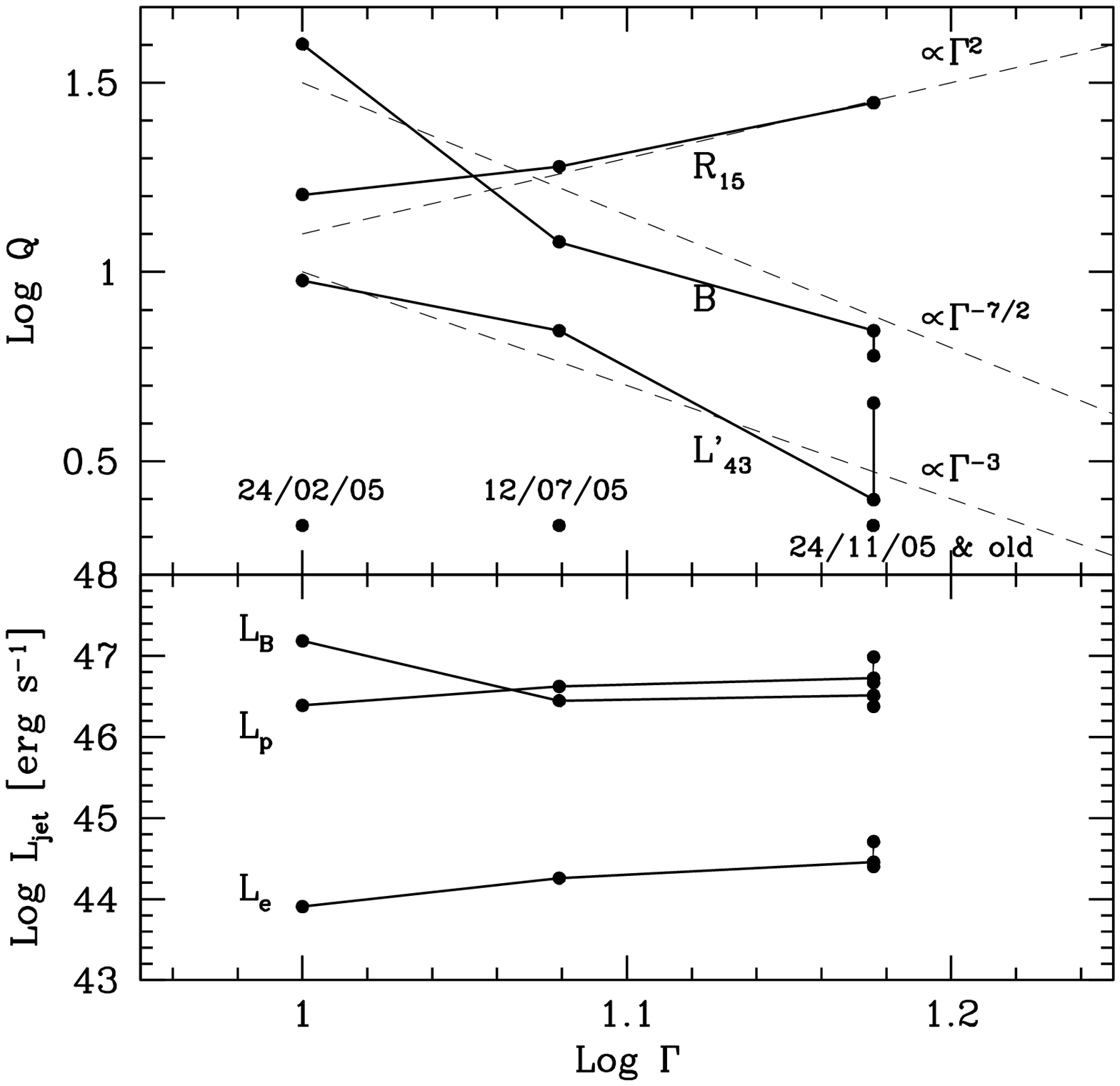,width=150mm,height=130mm,angle=0.0}
   \parbox{180mm}{{\vspace{2mm} }}
   \caption{Jet parameters of  PKS~0537-441. 
{\it Top panel:}  The logarithms of 3 quantities ("Q") are reported as 
a function of the logarithm of the bulk Lorentz factor:
the size of the emitting source $R_{15}$ in units of $10^{15}$ cm, the value of the 
magnetic field $B$ in Gauss, and the injected power 
$L^\prime_{43}$ 
(in the comoving frame) in the 
form of relativistic particles, in units of  $10^{43}$  erg~s$^{-1}$,  as used for our modelling.
The dashed lines represent  the relationships predicted by the Katarzy\'nski \& Ghisellini (2007) model.
The labelled dates identify the specific model/state of the source
(see  Pian et al. 2007 for more details on the model parameters). 
{\it Bottom panel:}
The power carried by the jet in the form of magnetic field 
($L_B$), cold protons ($L_{\rm p}$), relativistic electrons ($L_{\rm e}$)
resulting from our modelling, as a function of the bulk Lorentz factor (from Pian et al. 2007).}
   \end{center}
\end{figure}

\subsection{ \itbf{Swift} observations  of PKS~2155-304 following a giant TeV outburst}

PKS~2155-304 ($z = 0.116$)  is one of the extragalactic sources most frequently monitored by the current  experiments for the detection of Cerenkov  light induced by TeV energy radiation.  In July 2006 the blazar was detected by the HESS telescope at a level ten times higher than usual for this source.  On 28 July 2006 the TeV flux at energies larger than 200 GeV was 7 times larger than that of the Crab Nebula in the same energy interval (Aharonian et al. 2007).  This  triggered multiple instruments for follow-up observations at lower energies, including {\it Swift} (Foschini et al. 2007). 

A bright X-ray flare was detected by the {\it Swift} XRT about 1 day after the TeV outburst, which decreased thereafter by a factor of $\sim$5   in 1 month.  The X-ray spectral changes are not as dramatic.   In particular the frequency of the synchrotron peak remained at values 
similar to those observed in the past (e.g., 1997, Chiappetti et al. 1999), during low
TeV activity. Modeling of the spectral energy distribution (reported in Figure 5)  based on the synchrotron self-Compton process in a homogeneous 
region suggests  an increase of the Doppler factor ($33$ in $2006$; $18$ in $1997$) 
and of the normalization of the relativistic electrons distribution, associated with a decrease of the magnetic field
($0.27$ G in $2006$; $1$ G in $1997$, see Foschini et al. 2007).  This suggests that in this source, the observed variability cannot be solely reproduced with a variation of the bulk Lorentz factor, but other physical quantities must change between the observed states.


\begin{figure}
  \vspace{2mm}
   \begin{center}
   \hspace{3mm}\psfig{figure=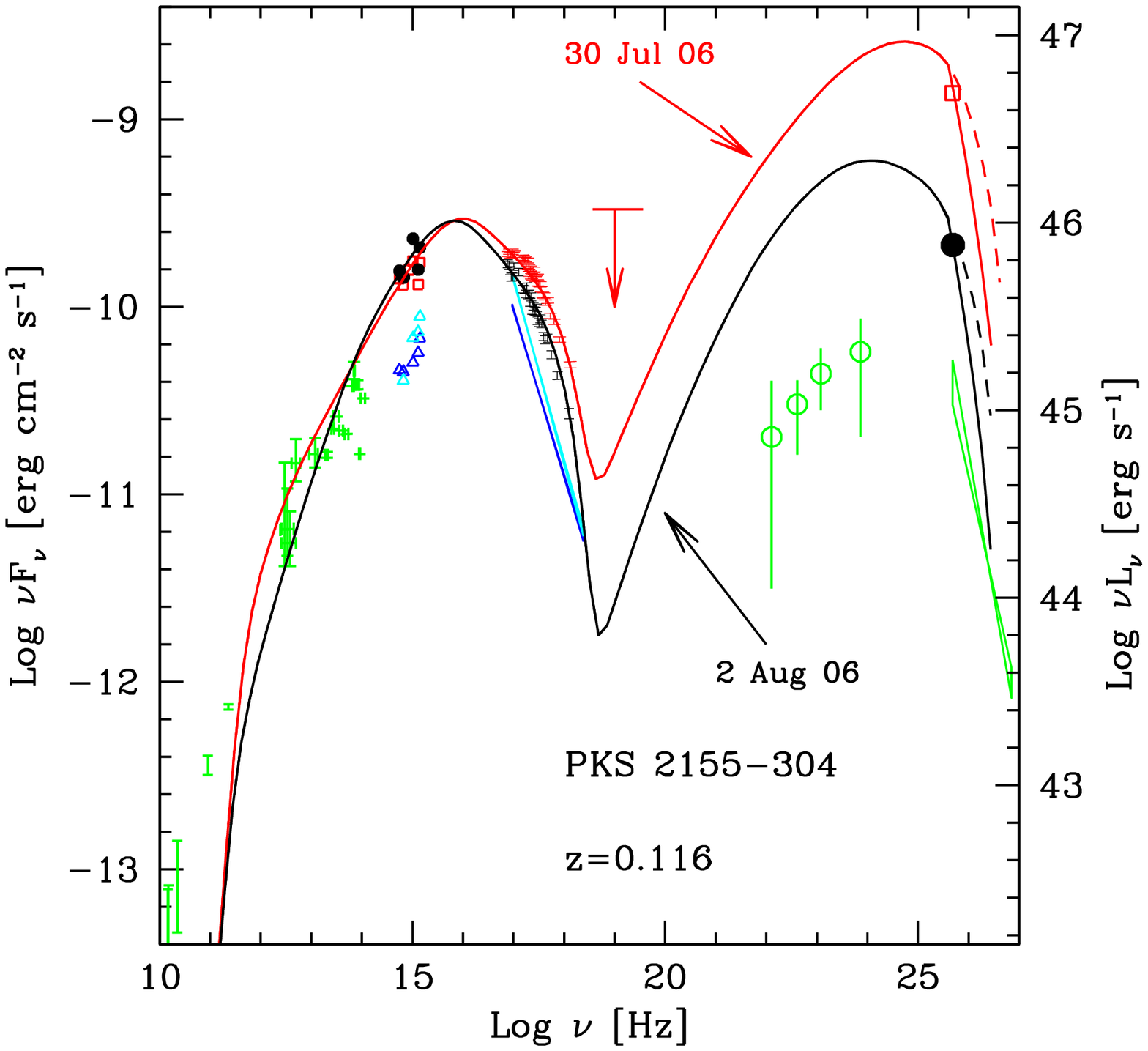,width=150mm,height=130mm,angle=0.0}
   \parbox{180mm}{{\vspace{2mm} }}
   \caption{Spectral energy distributions  of PKS~2155--304: the red
symbols represent the  quasi-simultaneous TeV (HESS) and X-ray ({\it Swift} XRT) data; 
the black symbols
refer to the TeV, XRT and REM observations of 2 August 2006 (see references in
Foschini et al. 2007).  For comparison, historical data are also shown:
green symbols refer to 1997 and previous epochs (see references in
Chiappetti et al. 1999) and to 2003 (HESS TeV spectrum taken in
October-November 2003, Aharonian et al. 2005), while in blue and light blue 
are reported the  \emph{XMM-Newton} data from Foschini et al. (2006). The red
and black continuous curves represent the synchrotron self Compton models 
(see Ghisellini et al. 2002) used to fit the data of July 2006 and August 2006, 
respectively.   Both models include the absorption at TeV energies due to the
extragalactic infrared background calculated according to Stecker \&
Scully (2006).  The dashed curves indicate the instrinsic (i.e. not
absorbed) spectrum (from Foschini et al. 2007).}
   \end{center}
\end{figure}

\section{Discussion}
\label{sect:discussion}

We have presented the  multiwavelength distributions of three well known and studied blazars at different epochs.  Two of the sources (3C~454.3 and PKS~0537-441)  have luminous BLRs and  therefore represent a benchmark for the economic jet model of Katarzy\'nski \& Ghisellini (2007) based on the internal shock scenario.   In this model, the flares are produced {\it within} the BLR, at different locations along the jet, from the collision of two consecutively emitted plasma blobs.  Depending on the distance of the dissipation site from the nucleus the plasma will move with different bulk Lorentz factors, larger values being attained farther from the nucleus.  Since the ratio between the external Compton 
and the synchrotron power in the blazar spectrum depends on the square of the bulk Lorentz factor (the external radiation field density, in the frame comoving with the blob, depends on $\Gamma^2$), the distance at which the blobs collide determines the relative importance of the two emission components and the shape of the overall spectrum.  Synchrotron-dominated multiwavelength blazar spectra are produced by collisions occurring closer to the jet apex, while the external Compton component, mainly responsible for the production of the MeV-GeV spectra, dominates when the flare is generated farther from the nucleus and closer to the BLR.    All physical quantities can be parameterized as functions of $\Gamma$, and their variations then depend on the changes of 
$\Gamma$.

For blazars with no luminous BLR, the economic jet model -- the concept of which is based on the  
relative distance of the dissipation site from the nucleus and from the BLR -- 
cannot be adequately tested.  While internal shocks can take place in these objects as well,  their observed multiwavelength variability must be explained with intrinsic changes of other physical quantities, beside $\Gamma$.  In the case of PKS~2155-304, these are the magnetic field and the electron distribution normalization (\S 3).   This implies a change in the total energy budget of the jet. The resulting variability affects the broad-band spectrum in a coherent way, producing a brightening
at all frequencies.

Parameter changes independent from $\Gamma$ can obviously take place also in objects with a rich BLR, but our purpose here is to demonstrate that this is not necessary: the very different observed multiwavelength states in these sources {\it can} be described by the dissipation of a fixed amount of energy at any given epoch.    
It must be noted that  our approach does not imply that the kinematics in the jets of blazars with and without luminous BLRs is {\it intrinsically} different: as said above, internal shocks can occur in both types of blazars.  However, a significant difference is apparent in the VLBI jet structures of EGRET blazars (typically exhibiting also prominent optical and UV emission lines) and BL Lac objects with no detected emission lines (Jorstad et al. 2001a; Piner \& Edwards 2004).

The key interesting features of the internal shock scenario is the relatively low
radiative efficiency (of order of a few per cent) which well accounts
for the dissipation of kinetic energy in blazars, as higher
dissipation rates would be difficult to reconcile with the large amount of power carried  up to the large scale lobes.
Furthermore, although the radiative dissipation occurs on
all jet scales most of it is localized within
tenths of a parsec, on the BLR scale, in agreement with the requirements
of fast variability and transparency to $\gamma$--rays (Spada et al. 2001).

Many VLBI campaigns have been organized with the aim of correlating with confidence the occurrence of a blazar multiwavelength outburst and the appearance of a new radio component in the jet (e.g., Jorstad et al. 2001b;
Savolainen et al. 2002; Lindfors et al. 2006).  
The velocities of the emerging plasma blobs may clarify better their behavior  within light-days from the nucleus, at scales that the VLBI cannot probe,  and help test the economic jet model, although the difficulties of disentangling the kinematical from the viewing angle effects may be unsurmountable.
We stress that the model has a high predictive power at the MeV-GeV energies (see e.g. Figure 3), so that the monitoring of blazars with {\it AGILE} and {\it GLAST}, coordinated with 
observations at lower energies, will represent a crucial test.

\begin{acknowledgements}
We would like to acknowledge the contribution of many colleagues to the success of the blazar observing campaigns described in this paper.  EP would like to thank the organizers of the Frascati Workshop 2007 for a very pleasant and stimulating conference.   This work has been supported by  the Italian MIUR and by the Italian Space Agency through the contract ASI-INAF I/023/05/0.
\end{acknowledgements}

\appendix                  

\section{DISCUSSION}

\noindent
{\bf J. Beall:}
Can you comment about the relationship between the Lorentz factor frequency shifts 
($\nu_{IC}  \propto \gamma^2 \nu_S$) and the bulk Lorentz factors between the synchrotron and self-Compton and the external Compton models? 

\bigskip

\noindent
{\bf E. Pian}: In the synchrotron self-Compton model the dependence between the maximum synchrotron frequency and the maximum inverse Compton frequency is mediated by the square of the random Lorentz factor of the particles.  In the external Compton  model the maximum frequency of the Compton component is still dependent on the shape of the incident electron population spectrum, but also depends on the spectral peak of the external photon field component.   

\bigskip

\noindent
{\bf S. Colafrancesco}
Blazars show probably different components (which are even variable) at  gamma-ray energies: some are associated with the core-dominated emission, others with the beamed jets.  Are the MeV-GeV
next experiments like AGILE/GLAST the best way to probe your model?

\bigskip

\noindent
{\bf E. Pian}:
The angular resolution of AGILE  and GLAST will not allow us to resolve different components, no matter how large is the viewing angle (this is usually   very small in blazars, anyway, less than few degrees).  As observed in M87, there may be different knots along the jet flaring at many frequencies, including gamma-rays.   The blazar nuclear gamma-ray emission  that we would like to model is produced in a jet volume confined to few light-days of the nucleus and is variable with high amplitude and short time scales. These characteristics should allow us to disentangle the nuclear component with fairly high confidence from other components produced at larger distances from the jet apex. Incidentally, the association of the dramatically variable TeV emission of  M87 (time scale of less than a day) with the jet knot resolved by  HST (some tenths of a parsec across)  represents a problem, unless relativistic beaming is implied, which is unlikely, considering the jet large angle orientation with respect to the observer.

\bigskip

\noindent
{\bf A. De Rujula}
Your model clearly fails to reproduce the blazar radio data.

\bigskip

\noindent
{\bf E. Pian}:
radio data usually map jet zones larger than those where the high energy radiation is emitted and are therefore not necessarily matched by our homogeneous one-zone model.

\label{lastpage}

\end{document}